\documentclass[preprintnumbers]{revtex4}
\usepackage{graphicx}
\def\CP{$ C \! P$ } 
\def\CPn{$ C \! P$} 

\begin{document}
\preprint{Snowmass P3-41, Bergen ISSN 0803-2696/2001-04}

\bibliographystyle{revtex}


\title{Experimental Status and Expectations Regarding Radiative 
Penguin Decays}



\author{G. Eigen}
\email[]{eigen@asfys2.fi.uib.no}
\affiliation{University of Bergen}


\date{\today}

\begin{abstract}
Radiative penguin decays provide a hunting ground complementary to direct 
searches for physics beyond the Standard Model. In the era of $B$-factories 
copious production of $B$ mesons permits precision measurements of radiative
penguin decays. We present herein the status of
radiative penguin processes and expectations at high luminosities,
focusing on $b \rightarrow s (d) \gamma$,  
$b \rightarrow s \ell^+ \ell^-$, and $b \rightarrow s \nu \bar \nu$ modes. 
\end{abstract}

\maketitle


%
%

%
%



\bibliographystyle{revtex}


\section{Introduction}
Radiative penguin decays are flavor-changing neutral current (FCNC) 
transitions that are forbidden in the Standard Model 
(SM) at tree level but occur at the loop level involving electroweak 
penguin loops or box diagrams. Though suppressed in SM they are relatively 
large in $b \rightarrow s$ because of the CKM structure and the 
top-quark dominating the loop. Additional contributions
can arise from New Physics effects such as new gauge bosons, 
charged Higgs bosons or supersymmetric particles. These interfere with the
SM processes. Depending on the sign of the interference term enhanced or 
depleted branching fractions result. In addition, due to the presence of 
new weak phases \CP asymmetries that are small in the SM may be enhanced. In 
this report we focus on electroweak penguin decays with a photon, a lepton
pair or a neutrino pair in the final state.
We have chosen five benchmark luminosities ${\cal L}$ for our extrapolations: 
(i) $9.1/20.7 \ \rm fb^{-1}$, integrated luminosity 
used in present analysis samples by CLEO and BABAR, 
(ii) 100 $\rm fb^{-1}$, integrated luminosity expected 
in BABAR by summer 2002, 
(iii) 500  $\rm fb^{-1}$, integrated luminosity expected in BABAR by
summer 2005,
(iv) $1 \ \rm ab^{-1}$, integrated luminosity
expected in BABAR by 2008, and 
(v) $10 \ \rm ab^{-1}$, annually integrated luminosity of a super $B$-factory
\cite{seeman}, \cite{burchat}.
We use the most precise measurements where available
and scale yields linearly with ${\cal L}$ and statistical errors 
by $1/\sqrt{\cal L}$. 
For modes that have not been observed yet we use a range of most recent 
predictions and inflate statistical errors by $\sqrt2$ to account 
conservatively for background subtraction. 
Systematic errors are a guess assuming that for 
increased data samples individual systematic uncertainties can be reduced,
by obtaining an improved understanding of the detector performance with time
and by choosing a set of analysis criteria that yield 
improved systematic errors even at a cost of reduced statistics.
Note that these estimates are intended as a guideline and need to be
backed up by detailed Monte Carlo studies. In particular, the 
systematic-error estimates need to be confirmed with detailed studies.


\section{Inclusive and Exclusive $b \rightarrow s (d) \gamma$ Modes}

\begin{figure}
\includegraphics[width=17cm]{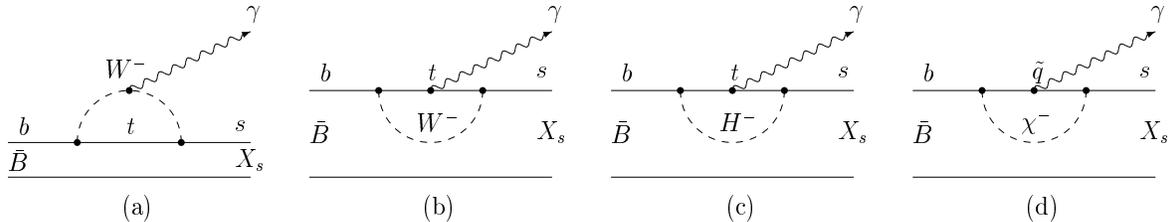}%
\caption{Lowest order Feynman diagrams for $b \rightarrow s \gamma$ decays in
the SM (a,b) and for New Physics contributions from a charged Higgs (c)
and supersymmetric processes (d).}
\label{fig:bsg}
\end{figure}

The electromagnetic penguin process $b \rightarrow s \gamma$ is dominated 
by the magnetic penguin operator $O_{7 \gamma}$. 
The SM decay rate
contains the squares of the CKM matrix elements $\mid V_{ts} \mid$ and the
Wilson coefficient $C_7$. The latter accounts for all perturbative QCD 
contributions. Due to operator mixing an effective coefficient results,
which in leading order (LO) takes the value
$C_7^{(0) eff} = -0.312^{+0.059}_{-0.034}$. Including the next order and 
employing a low-energy cut-off on the photon energy in the  
gluon-bremsstrahlung process 
yields an effective Wilson coefficient $\mid D^{eff} \mid = 0.373$.   
The non-perturbative contributions are absorbed into
the hadronic matrix element of the magnetic dipole operator. Because of
large model uncertainties one avoids the calculation of the hadronic matrix 
element by using the approximation that the ratio of decay rates of 
$b \rightarrow s \gamma$ and $b \rightarrow c e \bar \nu$ at the parton 
level is equal to that at the meson level.
New Physics processes yield additional contributions 
$C_7^{new}$ and $C_8^{new}$, where the latter arises from SUSY operators
that are equivalent to the chromomagnetic dipole operator $O_8$.  
Typical Feynman diagrams 
for SM and New Physics processes are shown in Figure~\ref{fig:bsg}. 
In next-to-leading order (NLO) the SM inclusive branching fraction is 
predicted to be ${\cal B}(B \rightarrow X_s \gamma) = (3.28 \pm 0.33) \times 
10^{-4}$ \cite{misiak1}. Gambino and Misiak \cite{misiak2}, however, have 
recently argued for a different choice of the charm-quark mass, which increases
the branching fraction to 
${\cal B}(B \rightarrow X_s \gamma) = (3.73 \pm 0.3) \times 10^{-4}$. 
The present theoretical uncertainty of $\sim 10\%$ is dominated by the 
mass ratio of the $c$-quark and $b$-quark and the choice of the scale 
parameter $\mu_b$.

So far inclusive measurements have been performed by CLEO \cite{cleo1}, 
BELLE \cite{belle1} and ALEPH \cite{aleph1}, of which the CLEO result is the 
most precise. The analysis is an update extending the observed photon-energy 
range to 2.0- 2.7~GeV ($94\%$ of the spectrum). The main 
backgrounds originate from $q \bar q$ continuum processes with either a 
high-energy photon from initial-state radiation (ISR) or from a $\pi^0$. 
To reduce these backgrounds, CLEO exploits several event-shape 
variables, performs $B$-meson pseudoreconstruction and uses kinematic
information of identified leptons. Candidates are sorted into four classes:
events selected solely with event-shape variables, those having in addition
a $B$ pseudoreconstruction, those with an additional lepton, and
those satisfying all three requirements. 
In each class all variables are combined in a 
neural net, which computes a weight between 0.0 and 1.0, depending on how much 
the event is continuum-like or $B \rightarrow X_s \gamma$ signal-like. 
The observed spectrum containing $1861.7 \pm 16.5$ weights in the signal 
region is still dominated by backgrounds ($75\%$ continuum, 
$12.3\%$ $B \bar B$). 
After background subtraction, where the continuum-background spectrum 
is obtained from data taken below the $\Upsilon(4S)$ resonance and 
the $B$-background spectrum is determined from $B \bar B$  
Monte Carlo, which was tuned to match yields observed in the data, 
CLEO finds a $B \rightarrow X_s \gamma$ signal
yield of $ 233.6 \pm 31.2\pm 13.4$ weights in a sample of $9.1 \ \rm fb^{-1}$. 
With a detection efficiency of $\epsilon = (3.93 \pm 0.15 \pm 0.17)\%$ 
CLEO measures a branching fraction of ${\cal B}(B \rightarrow X_s \gamma) =
(3.21 \pm 0.43_{(stat)} \pm 0.27_{(sys)}$$
^{+0.18}_{-0.10\ (th)}) \times 10^{-4}$, 
where errors are statistical, systematic and from theory,
respectively. This result is consistent with the SM prediction and
agrees with the BELLE measurement of ${\cal B}(B \rightarrow X_s \gamma) =
(3.36 \pm 0.53_{(stat)} \pm 0.42_{(sys)}$$^{+0.50}_{-0.54\ (th)}) 
\times 10^{-4}$. 

Presently, errors are rather large amounting to a relative statistical 
(systematic) error of $13.4\% \ (8.4\%)$. They are slightly larger than the
present theoretical uncertainty. 
Using the CLEO measurement, the yields and relative errors obtained from 
extrapolations to high ${\cal L}$ are summarized in 
Table~\ref{tab:bsg}. Note that a super $B$-factory 
operating at a luminosity of $\rm 10^{36} \ cm^{-2} s^{-1}$
is expected to produce $2.6 \times 10^5$ $B \rightarrow X_s \gamma$
signal weights per year permitting a $B \rightarrow X_s \gamma$
branching-fraction measurement with 
a relative statistical error of $0.4\%$. 
It is expected that with increased statistics the systematic error can be 
reduced substantially by using appropriate data selections even at the cost
of slightly reduced statistics and by improving
measurements of tracking efficiency, photon energy, photon efficiency
and $B$ counting.
The precision of the SM prediction needs to be improved to ascertain a high 
sensitivity for New Physics processes. In a hadron collider precision
measurements are difficult because of high backgrounds.

\begin{table} [hbtp] \centering
\caption [ ] {Yields $Y_{\cal B} \ (Y_{CP})$, 
statistical errors $\sigma_{stat}/{\cal B} \ (\sigma_{stat}^{CP})$
and systematic errors $\sigma_{sys}/{\cal B} \ (\sigma_{sys}^{CP})$
expected for branching-fraction (\CPn -asymmetry) measurements 
of $B \rightarrow X_s \gamma$ and $B^0 \rightarrow K^{*0} \gamma$
for different luminosities.}
\label{tab:bsg}
\medskip
\begin{tabular} { | l ||c|c|c|c|c|c| }  \hline 

${\cal L} \ \rm[fb^{-1}/y]$ & 9.1  & 20.7 & 100 & 500 & 1000 & 10000 \\ 
\hline \hline
$X_s \gamma$ weights $Y_{\cal B} \ (Y_{CP})$ & 234 (231) & 
& 2570 \ (2540) &  $1.28 \ (1.27)\times 10^4  
$ &  $2.57 \ (2.54)\times 10^4 $
& $2.57 \ (2.54) \times 10^5$ \\ \hline
$\sigma_{stat}/{\cal B} \ (\sigma_{stat}^{CP})\ [\%]$ 
& 13.4 (10.8) & & 4.0 (3.3)& 1.8 (1.5)& 1.3 (1.0)
& 0.4 (0.33) \\ \hline
$\sigma_{sys}/{\cal B} \ (\sigma_{sys}^{CP})\ [\%]$ 
& 8.4 (2.2)& & 5 (1.8)& 3 (1.5)& 2 (1.0)& 1-2 (0.5)\\ \hline \hline
$K^{*0} \gamma$ yield  $Y_{\cal B} \ (Y_{CP})$& 
& 139.2 (139.2) & 670 (670) & 3360 (3360)& $ 6.72 \ (6.72) \times 10^3$& 
$6.72 \ (6.72) \times 10^4$ \\ \hline
$\sigma_{stat}/{\cal B}  \ (\sigma_{stat}^{CP})\ [\%]$ 
& & 9.3 (9.4) & 4.2 (4.3)& 1.9 (1.9)& 1.3 (1.4)& 
0.42 (0.43) \\ \hline
$\sigma_{sys}/{\cal B}  \ (\sigma_{sys}^{CP})\ [\%]$ 
& & 6.2 (1.2) & 4 (1.0)& 3 (0.8)& 2 (0.7)
& 1? (0.5) \\ \hline 
\end{tabular} 
\end{table}

\begin{figure}
\includegraphics[width=10cm]{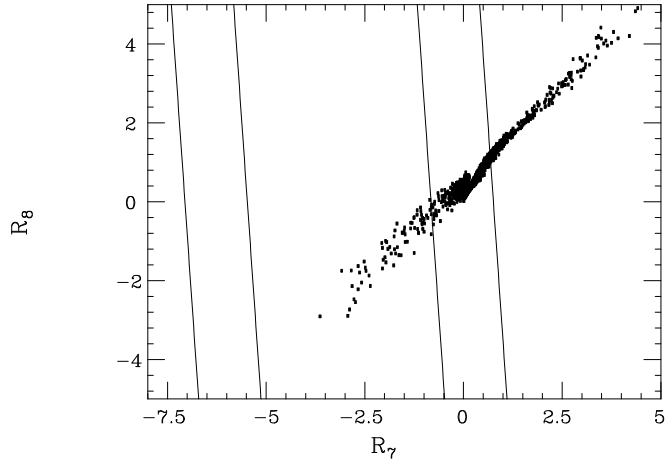}%
\caption{Scatter plot of $R_8$ versus $R_7$ for solutions obtained
in the SUGRA model. The region allowed by the CLEO measurement lies 
inside the two sets of solid diagonal bands. 
\label{fig:susy}} 
\end{figure}

The present ${\cal B} ( B \rightarrow X_s \gamma)$ measurements 
already provide a significant constraint on the SUSY parameter space. 
For example the new physics contributions to $B \rightarrow X_s \gamma$, 
$C_7^{new}$ and $C_8^{new}$, have been calculated using the minimal 
supergravity model (SUGRA) \cite{hewett1}. Many solutions have been
generated by varying the input parameters within the ranges 
$ 0 < m_0 < 500$~GeV, $50 < m_{1/2} < 250$~GeV, $ -3 < A_0/m_0 < 3$ and 
$ 2 < \tan \beta < 50$ \footnote{In SUGRA one assumes a
common scalar mass $m_0$ for squarks and sleptons, a common gaugino mass
$m_{1/2}$ and a common trilinear scalar coupling $A_0$. As usual the ratio
of vacuum expectation values of the neutral components of the two Higgs 
doublets is parameterized by $\tan \beta$.},
while the top-quark mass was kept fixed at
$ m_t = 175$~GeV. Only solutions were retained that were 
not in violation with  SLC/LEP
constraints and Tevatron direct sparticle production limits. For these
the ratios $R_7= C_7^{new}(M_W)/C^{SM}_7(M_W)$ and 
$R_8= C_8^{new}(M_W)/C^{SM}_8(M_W)$ were determined. The results are depicted 
in Figure~\ref{fig:susy} \cite{hewett2}. The solid bands show the regions 
allowed by the CLEO measurement. It is interesting to note that many 
solutions are already in conflict with the data.

The exclusive decay rate for $B \rightarrow K^* \gamma$ involves 
the hadronic matrix of the magnetic dipole operator, which in general
is expressed in terms of three $q^2$-dependent form factors $T_i(q^2)$. 
For on-shell photons $T_3$ vanishes and $T_2$ is related to $T_1$. For the
determination of the form factors various techniques are used, introducing
additional theoretical uncertainties.
Recently, two NLO calculations were carried out, predicting SM branching 
fractions of 
${\cal B} (B  \rightarrow K^* \gamma) = (7.1^{+2.5}_{-2.3}) \times 10^{-5}$
\cite{bosch} and 
${\cal B} (B  \rightarrow K^* \gamma) = (7.9^{+3.5}_{-3.0}) \times 10^{-5}$
\cite{beneke}. 
The exclusive $B \rightarrow K^* \gamma$ modes have been studied by 
BABAR \cite{babar1}, BELLE \cite{belle2} and CLEO \cite{cleo2}, where
BABAR used the highest statistics sample. 
Utilizing kinematic constraints resulting from a full $B$ 
reconstruction in the $B$ rest frame provides
a substantial reduction of the $q \bar q$-continuum background here.
We base our extrapolations to high ${\cal L}$ on the BABAR 
$B^0 \rightarrow K^{*0} \gamma$ result in the $K^+ \pi^-$ final state, 
where a reconstruction efficiency of $14\%$ is achieved. 
In a sample of ${\cal L} = 20.7 \ \rm fb^{-1}$ a yield of 
$139.2 \pm 13.1$ events is observed, resulting in a branching fraction 
of ${\cal B}(B^0 \rightarrow K^{*0} \gamma) = 
(4.39 \pm 0.41_{(stat)} \pm 0.27_{(sys)}) \times 10^{-5}$. 
Due to the large theoretical errors of $35-40\%$ the BABAR measurement
is still consistent with the NLO SM predictions. Note that the combined
statistical and systematic error is already more than a factor of three 
smaller than the theoretical uncertainty. The results of
our extrapolations to high ${\cal L}$ are also shown in 
Table~\ref{tab:bsg}.  
Expected precisions are similar to those in $B \rightarrow X_s \gamma$.
In hadron colliders $B \rightarrow K^{*0} \gamma$ is also measurable. CDF 
expects to observe $170 \pm 40$ events per 
$2 \ \rm fb^{-1}$, while BTEV \cite{tev}
and LHCb \cite{lhc} estimate yields of
27000 and 26000 events per $10^7 s \ (\sim 2\ \rm fb^{-1})$, 
respectively.

\CP asymmetries provide another test of the SM. While small in the SM 
($\leq 1\%$) \cite{soares}
they may be as large as $20\%$ \cite{kagan} in SUSY models. So far all
observed \CP asymmetries are consistent with zero. In the inclusive mode
we base our extrapolations on a recent result from CLEO \cite{cleo3}, yielding
${\cal A}_{CP}(B \rightarrow X_s \gamma) = 
(-0.079 \pm 0.108 \pm 0.022) \times (1.0
\pm 0.03)$. The first error is statistical, while 
the second and third errors represent additive and multiplicative systematic 
uncertainties, respectively. For extrapolating \CP asymmetries of the 
exclusive $B^0 \rightarrow K^{*0} \gamma$ modes to high ${\cal L}$,  
we use the BABAR result of ${\cal A}_{CP}(B \rightarrow K^{*0} \gamma) 
= -0.035 \pm 0.094_{(stat)} \pm 0.012_{(sys)}$ obtained in the 
$K^+ \pi^-$ final state \cite{babar1}. The extrapolated yields and errors
are listed in Table~\ref{tab:bsg} in parentheses.
Adding the
$K^+ \pi^0$ and $K_S^0 \pi^+$ final states increases the yield to
$225.2 \pm 17.9$ events. The asymmetry remains unchanged, just
the statistical error is reduced to $7.6\%$. 
 While New Physics at 
the $20\%$ level should be visible in BABAR by next summer, a super 
$B$-factory is needed to uncover New Physics at the few $\%$ level.

Both inclusive and exclusive $b \rightarrow d \gamma$ decays, which are 
suppressed by $\mid V_{td} / V_{ts} \mid^2$ with respect to corresponding
$b \rightarrow s \gamma$ modes, have not been seen yet. 
A branching-fraction measurement of 
$B \rightarrow X_d \gamma$ provides a determination
of $\mid V_{td} / V_{ts} \mid$ with small theoretical uncertainties.
However, backgrounds are expected to be huge, since this mode is
CKM-suppressed and $u \bar u, d \bar d$ continuum processes are enhanced
compared to $s \bar s$ continuum processes. An NLO calculation, 
which includes long-distance effects of $u$ quarks in the penguin loop,
predicts a range of
$ 6.0 \times 10^{-6} \leq {\cal B}(B \rightarrow X_d \gamma) \leq 2.6 \times
10^{-5}$ \cite{ali2} for the inclusive branching fraction. 
The uncertainty is dominated by imprecisely known CKM parameters. 
Due to the enormous backgrounds a full or at least partial reconstruction 
of the other B-meson is probably needed. Using the above range of
branching-fraction predictions and assuming a reconstruction efficiency of 
$0.1\%$ we estimate luminosities in the range of 
${\cal L} = 20-4.7\ \rm ab^{-1}$ to achieve a $6.5 \ \%$ statistical accuracy 
on $\mid V_{td} / V_{ts} \mid$, thus requiring 2-0.5 years of running
at a super $B$ factory.
A determination of $\mid V_{td} / V_{ts} \mid$ in the exclusive modes
$B \rightarrow \rho(\omega) \gamma$ bears enhanced model uncertainties,
since form factors are not precisely known. The branching fraction 
for $B \rightarrow \rho(\omega) \gamma$ is reduced by a factor 
$\sim 20$ with respect to 
$B \rightarrow K^* \gamma$. In addition, long-distance effects
may increase branching fractions by a factor of two \cite{bbrpb}. For 
${\cal B} (B \rightarrow \rho  \gamma)/{\cal B} (B \rightarrow K^*  \gamma) 
= 0.05$ and an efficiency of $7\%$ we would need 
$ {\cal L}= 0.72 \ (18) \ \rm ab^{-1}$
to obtain a $10 \ (2) \%$ statistical accuracy in the branching fraction.
The \CP asymmetry predicted in SM for $B \rightarrow \rho \gamma$ is of
the order of $10\%$ \cite{greub}.

\section{Inclusive and exclusive $b \rightarrow s \ell^+ \ell^-$ Modes}

\begin{figure}
\includegraphics[width=17cm]{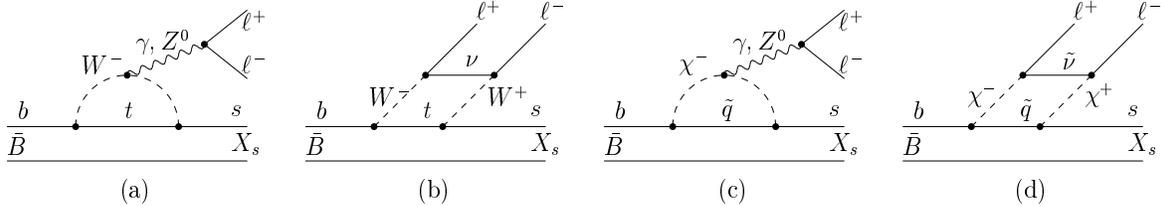}%
\caption{Feynman diagrams for $b \rightarrow s \ell^+ \ell^-$ decay in 
the SM  (a,b), and for supersymmetry contributions (c,d).}
\label{fig:bsll}
\end{figure}

The radiative decays $b \rightarrow s \ell^+ \ell^-$ are suppressed with
respect to  $b \rightarrow s \gamma$ by about two orders of magnitude.
The suppression by $\alpha$ is compensated partially by additional 
contributions from the $Z^0$-penguin diagram and a box diagram that involves
the semileptonic operators, $O_{9V}$ and $O_{10A}$. Each of them can receive 
additional SUSY contributions. Characteristic Feynman diagrams 
are depicted in Figure~\ref{fig:bsll}. New Physics processes may enhance or 
deplete decay rates with respect to predictions in SM. 
Models are characterized in terms of ratios of Wilson coefficients 
$R_i =1 +C_i^{NP}/C_i^{SM}$ for $i=7,9,10$.
As an example Figure~\ref{fig:bkll} depicts the dilepton-mass-squared spectrum
for $B \rightarrow K^* \mu^+ \mu^-$ calculated in SM, SUGRA models 
and minimal-insertion-approach SUSY models (MIA) \cite{lcsm}. 
The SM prediction is the lowest. However, due to form-factor related 
uncertainties it may be difficult to uncover
New Physics effects unless they are huge. 
It is interesting to point 
out that due to interference effects between the penguin process and the
long distance processes $B \rightarrow \psi(nS) K^*$ an enhanced (depleted)
rate is observed below (above) each $\psi(nS)$ resonance. This in fact may be
a useful tool to extract the penguin contribution from an observed 
dilepton-mass-squared spectrum.

For inclusive modes the SM predicts in NLO branching-fractions of
${\cal B}(B \rightarrow X_S e^+ e^-) = 
(6.3^{+1.0}_{-0.9}) \times 10^{-6}$ and  
${\cal B}(B \rightarrow X_S \mu^+ \mu^-) = 
(5.7 \pm 0.8)\times 10^{-6}$ \cite{ali3}, \cite{misiak3}, \cite{buras1}. 
So far only CLEO \cite{cleo4} has searched for 
$B \rightarrow X_s \ell^+ \ell^-$ setting branching-fraction 
upper limits that are almost an order of magnitude above the SM predictions. 
For our extrapolations shown in Table~\ref{tab:bsll}, 
we use the range of the SM predictions and efficiencies measured by CLEO of 
$\epsilon(X_s e^+ e^-) = 5.2\%$ and $\epsilon (X_s \mu^+ \mu^-) = 4.5\%$.
We have assumed a 1.1~nb $b \bar b$ cross section and an equal amount
of $B^0$ and $B^+$ production. High luminosities are required to
accumulate a reasonably large sample, thus emphasizing the need for a 
super $B$-factory. At hadron machines also large $X_s \ell^+ \ell^-$ samples 
are produced. The main issue, however, is whether backgrounds can be 
reduced sufficiently to make competitive measurements.

{\textwidth 10cm
\begin{figure}
\includegraphics[width=7 cm]{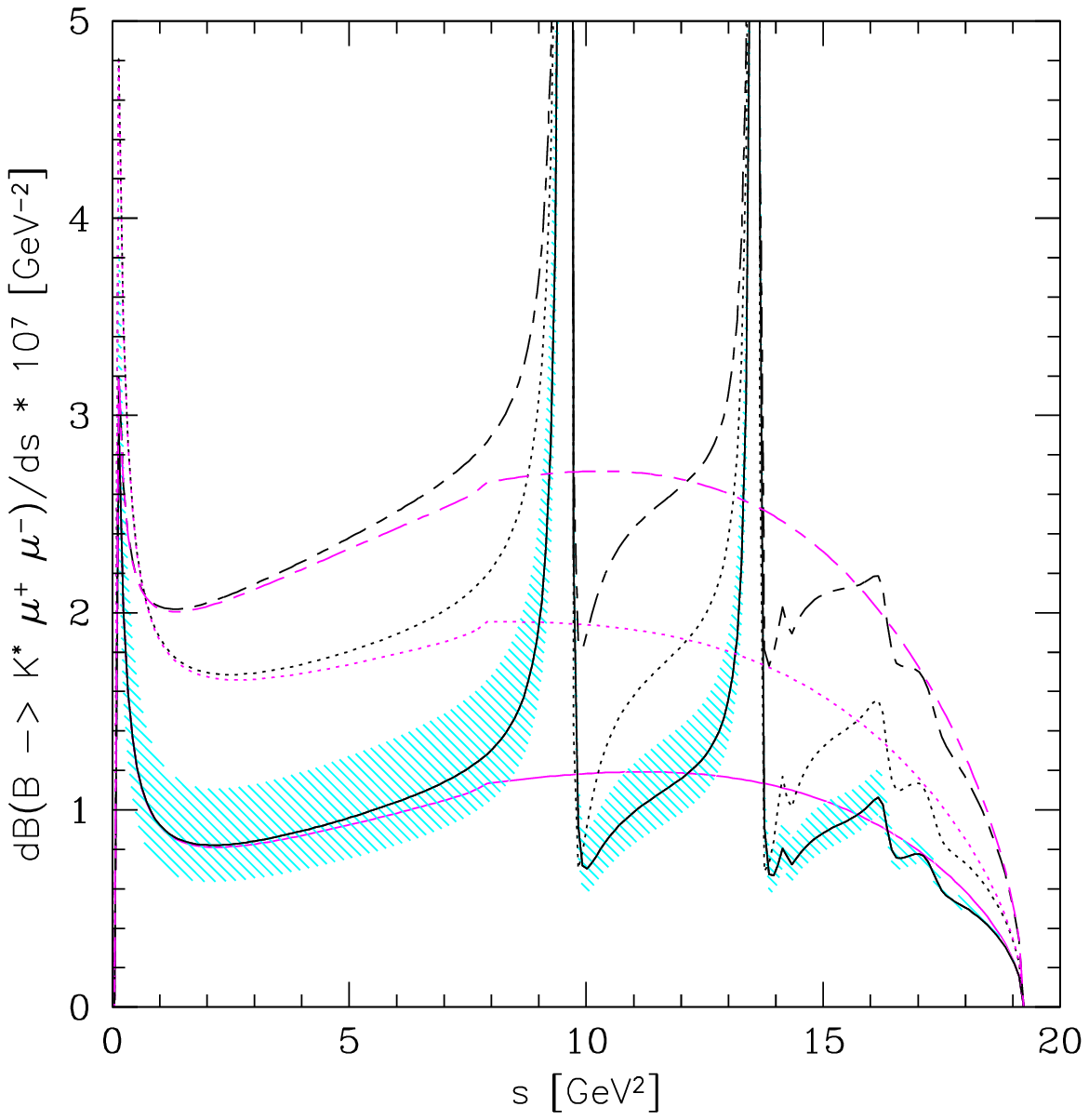}%
\includegraphics[width=7 cm]{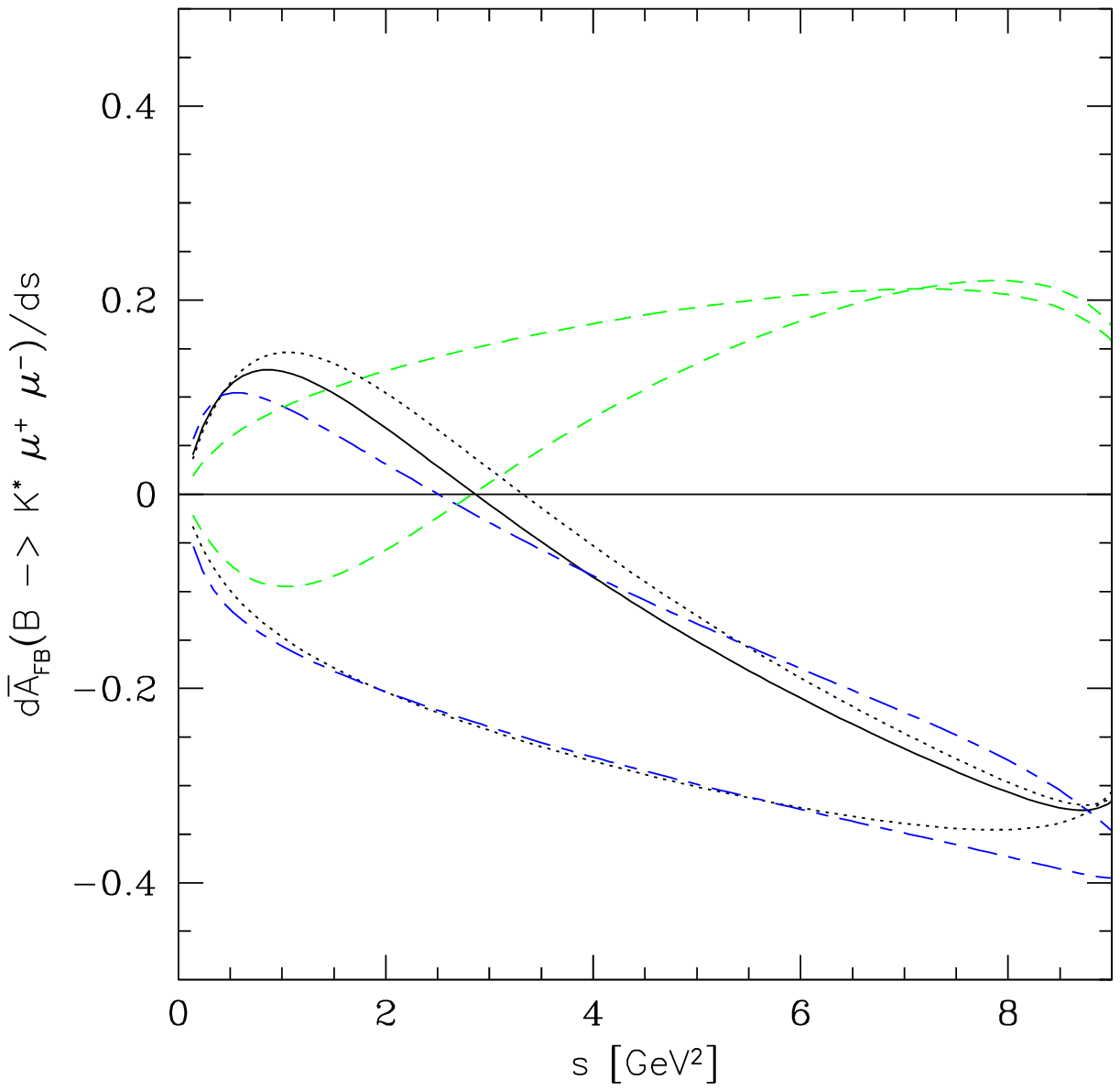}%
\caption{The dilepton invariant mass-squared spectrum (left) and the normalized
forward-backward asymmetry (right) as a function of $s = m^2_{\mu\mu}$ in
$B \rightarrow K^* \mu^+ \mu^-$ \cite{lcsm}. The solid lines denote
the SM prediction. The shaded region depicts form-factor related 
uncertainties. The dotted lines correspond to a SUGRA model ($R_7 =-1.2,
R_9=1.03, R_{10}=1$) and the dash-dotted lines to a MIA
model ($R_7 =-0.83,R_9=0.92, R_{10}=1.61$). In the $m^2_{\mu\mu}$
spectrum both the pure penguin contribution and the distribution including
long-distance effects are shown. In the ${\cal A}_{fb}$ plot
the upper and lower sets
of curves show the difference between $C^{(0)eff}_7 < 0$ and 
$C^{(0)eff}_7 > 0$, while the dashed curves give results for
another MIA model ($R_7 =\mp 0.83,R_9=0.79, R_{10}=-0.38$).
}
\label{fig:bkll}
\end{figure}
}

\begin{table} [hbtp] \centering
\caption [ ] {Event yields and relative statistical and 
relative systematic errors of branching fractions in 
$b \rightarrow s e^+ e^- \ (b \rightarrow s \mu^+ \mu^-)$ 
modes expected for different luminosities. 
Statistical errors include a factor of $\sqrt2$ to account for background 
subtraction. Systematic errors are guesses based on CLEO and BABAR.
}
\label{tab:bsll}
\medskip
\begin{tabular} { | l ||c|c|c|c|c|c|c| }  \hline 

${\cal L} \ \rm [fb^{-1}/y]$ & 20 & 100 & 500 & 1000 & 10000 \\ \hline \hline
$X_s \ell^+ \ell^- $ Yield & 
12-17 (10-13)& 62-84 (49-64)& 310-420 (240-320) 
& 620-835 (490-640) & 6180-8350 (4850-6440)\\ \hline
$\sigma_{stat}/{\cal B} \ [\%]$ & 40-35 (45-39)&18.0-15.5  (20-17.6)& 
8.0-6.9 (9.1-7.9)& 5.7-4.9  (6.4-5.6)& 1.8-1.5 (2.0-1.8)
\\ \hline
$\sigma_{sys}/{\cal B} \ [\%]$ &15 (25)& 10 (17)& 7 (12)& 6 (10)& 4 (7)? 
\\ \hline \hline
$K^+ \ell^+ \ell^-$ Yield & 
1.8-2.9  (1.1-1.7)& 9-14  (5-9)& 
45-72  (27-43) & 90-144 (54-87) & 905-1440 (540-870)\\ \hline
$\sigma_{stat}/{\cal B} \ [\%]$ 
&105-83 (136-107) & 47-37 (61-48) & 21-17 (27-21)
&14.9-11.8 (19.2-15.2) & 4.7-3.7 (6.1-4.8)  \\ \hline
$\sigma_{sys}/{\cal B} \ [\%]$ 
&14 (15) & 10 (12) & 8 (10) & 6 (7) & 3-4 (4-5)  \\ \hline \hline
$K^{*0} \ell^+ \ell^- $ Yield & 
3.1-6.7 (1.6-4.2)& 16-34 (8-21)& 80-170 (40-106)& 
160-340 (80-210)& 1570-3370( 790-2110) \\ \hline
$\sigma_{stat}/{\cal B} \ [\%]$ 
&80-55 (112-69) &36-24 (50-31) & 16.0-10.9 (22.5-13.8) &
11.3-7.7 (15.9-9.7)  & 3.6-2.4 (5.0-3.1)
\\ \hline
$\sigma_{sys}/{\cal B} \ [\%]$ & 14 (15)& 10 (12) & 7 (9) & 5 (7) & 3 (4)\\ \hline
\end{tabular} 
\end{table}

Branching fractions of the exclusive modes are further suppressed. 
Using predictions from a quark model \cite{qm} and light cone sum 
rules \cite{lcsm} we obtain the following ranges of SM predictions:
${\cal B}(B \rightarrow K \ell^+ \ell^-) = (4.7-7.5) \times 10^{-7}$,
${\cal B}(B \rightarrow K^* e^+ e^-) = (1.4-3.0) \times 10^{-6}$, and
${\cal B}(B \rightarrow K^* \mu^+ \mu^-) = (0.9-2.4) \times 10^{-6}$. 
BABAR \cite{babar1}, BELLE \cite{belle2} and CLEO \cite{cleo5}
have performed studies of the exclusive modes. 
Except for an unconfirmed signal seen by BELLE in the $K \mu^+ \mu^-$
final state with ${\cal B}(B  \rightarrow K \mu^+ \mu^-) =
(0.99 ^{+0.40}_{-0.32\ (stat)}$$^{+0.13}_{-0.14\ (sys)}) 
\times 10^{-6} $ \cite{belle3}, 
which is barely consistent with the BABAR limits,
no other signals have been observed yet. 
Using ${\cal L} = 20.7 \ \rm fb^{-1}$ BABAR has obtained the lowest $90\%$ 
CL branching-fraction upper limits:
${\cal B}(B \rightarrow K \ell^+ \ell^-) < 0.6 \times 10^{-6}$,
${\cal B}(B \rightarrow K^{*0} e^+ e^-) < 5.0 \times 10^{-6}$, and
${\cal B}(B \rightarrow K^{*0} \mu^+ \mu^-) < 3.6 \times 10^{-6}$.
While the $B \rightarrow K \ell^+ \ell^-$ branching fraction upper limit lies
amidst the SM predictions, the
$B \rightarrow K^{*0} e^+ e^- ( K^{*0} \mu^+ \mu^-)$ 
branching fraction upper limits are
less than a factor of two above the SM predictions.
For our extrapolations presented in Table~\ref{tab:bsll} 
we use efficiencies measured in BABAR: 
$\epsilon(K^+ e e) = 17.5\%$, $\epsilon (K^+ \mu \mu) = 10.5\%$,
$\epsilon(K^{*0} e e)= 10.2\%$ and $\epsilon (K^{*0} \mu \mu) = 8.0\%$.
At the Tevatron CDF and D0 expect to observe the
$K^{*0} \mu^+ \mu^-$ final state in a sample of $2 \ \rm fb^{-1}$ \cite{tev}.
Yield estimates are of the order of $59$ events for CDF and  
$\sim 310-130$ events for D0 depending on the lepton momentum requirement,
where D0 makes more optimistic assumptions than CDF. BTEV expects 
$K^{*0} \mu^+ \mu^-$ signal yields of 2240 events for $2 \ \rm fb^{-1}$ 
\cite{tev}. At LEP ATLAS, CMS and LHCb expect 
to observe 665, 4200 and 4500 events per $ 10^7 \ \rm s \ (\sim
2 \ fb^{-1})$, respectively \cite{lhc}.

The lepton forward-backward asymmetry ${\cal A}_{fb}(s)$ as a function of 
$s = m_{\ell \ell}^2$ is an observable that is very sensitive to SUSY
contributions. It reveals characteristic shapes in the SM both for
inclusive and exclusive final states. With sufficient 
statistics this asymmetry is a powerful tool to discriminate between SM and 
New Physics. To avoid complications from the
$\psi$ resonances one restricts the range to masses below the
$J/\psi$, which accounts for $\sim 40\%$ of the entire
spectrum. Figure~\ref{fig:bkll} shows ${\cal A}_{fb}(q^2)$ 
for the $B \rightarrow K^{*0} \mu^+ \mu^-$ mode \cite{lcsm}.
In SM the position $s_0$ of ${\cal A}_{fb}(s_0)=0$ is predicted to lie at 
$s_0 = 2.88 ^{+0.44}_{-0.28}\ \rm GeV^2$. Both, the shape and $s_0$
are expected to differ significantly in New Physics models. The shape is
very sensitive to the sign of $R_7$ and varies from model to model.
Thus, a precise measurement of  ${\cal A}_{fb}(q^2)$ may permit an
extraction of the coefficients $R_i$. 
The extrapolated yields in Table~\ref{tab:bsll} indicate
that a yearly luminosity of $10 \ \rm ab^{-1}$ is needed 
to determine ${\cal A}_{fb}(s)$ with reasonable 
precision. For Measuring 18 data points below $s = 9 \ \rm GeV^2$ with 
100~events each in the $B \rightarrow X_s \ell^+ \ell^-$
($B \rightarrow K^{*0} \ell^+ \ell^-$) modes at a super $B$ factory
($10 \ \rm ab^{-1}/y$)
requires a run period of 0.3-0.4 (0.8-1.3) years.

\section{Inclusive and Exclusive $b \rightarrow s \nu \bar \nu$ Modes}

The processes $b \rightarrow s \nu \bar \nu$ result from the $Z^0$ 
penguin or box diagrams by replacing $\ell^+, \ \ell^-, \ \nu$ 
in Figure~\ref{fig:bsll} with $\nu, \ \bar \nu, \ \ell^+ $, respectively. 
The branching fraction predictions are expected to bear  
the smallest model dependence among all radiative penguin decays,
since long distance effects are absent and QCD corrections are small.
The largest error results from the uncertainty of the $t$-quark mass.
Thus, these modes have the highest sensitivity to search for New Physics
contributions.  
The inclusive branching fraction in SM is predicted
to be ${\cal B}(B \rightarrow X_s \nu \bar \nu) = (4.1 ^{+0.8}_{-1.0}) 
\times 10^{-5}$ \cite{gln}, \cite{gln2}. The branching fractions for 
exclusive modes 
lie in the range ${\cal B}(B \rightarrow K \nu \bar \nu) = (2.4-9.2) 
\times 10^{-6}$ and ${\cal B}(B \rightarrow K^* \nu \bar \nu) = 
(0.8-2.6) \times 10^{-5}$ \cite{ali1}, \cite{qm}. 
So far, none of these modes has been observed. 
ALEPH has set a limit of ${\cal B}(B \rightarrow X_s \nu \bar \nu) 
< 7.7 \times 10^{-4} @ 90\% \rm CL$ , which is more than an
order of magnitude above the SM prediction. Due to the two 
unobserved neutrinos
one strategy of controlling backgrounds consists of a full reconstruction of 
the other $B$ meson. Presently,
the efficiency of fully-reconstructed $B$'s is $\sim 0.075\%$. 
Assuming that this can be increased by a factor of two by partial
reconstruction, and assuming detection efficiencies of
$80\%/43\%/33\%$ for $K^+/K^\pm \pi^\mp/X_s$ reconstruction, we obtain the 
extrapolated yields shown in Table~\ref{tab:bsnunu}. Since these modes are not 
accessible in hadron machines, a super $B$-factory is needed to observe them
and measure their properties. In a sample of  $\rm 10 \ ab^{-1}$
the statistical error for the $X_s \nu \bar \nu$ final state
in the optimistic case is still $6\%$.

\begin{table} [hbtp] \centering
\caption [ ] {Expected event yields for
$B \rightarrow X_S \nu \bar \nu^-$  and $B \rightarrow K^{(*)} \nu \bar \nu^-$
modes for different luminosities.}
\label{tab:bsnunu}
\medskip
\begin{tabular} { | l ||c||c|c|c|c|c| }  \hline 

${\cal L} \ \rm [fb^{-1}/y]$ & efficiency [$10^{-4}$]& 20  & 100 & 500 & 1000 & 10000 \\ 
\hline \hline
$K^+ \nu \bar \nu$ Yield & 12.0& 0.06-0.24& 0.3-1.2 & 1.6-6.1 & 3.2-12& 32-120 \\ \hline
$K^{*0} \nu \bar \nu$ Yield & 6.5& 0.1-0.4& 0.6-1.9 & 2.9-9.3 & 5.7-19& 57-186
 \\ \hline
$X_s \nu \bar \nu$ Yield & 5.0& 0.7-1.1 & 3.4-5.4 & 17-27 & 34-54& 340-540
 \\ \hline
\end{tabular} 
\end{table}

\section{Conclusion}

Present asymmetric $B$-factories will accumulate sufficient luminosities to
achieve precise branching-fraction and \CPn -asymmetry measurements in 
inclusive and exclusive $b \rightarrow s  \gamma$ decays 
allowing searches for physics beyond the SM. These measurements
are complementary to direct searches and may
yield positive results before the start of the LHC. 
The data samples in present asymmetric $B$ factories
will be sufficiently large to allow for a discovery
of inclusive and exclusive $b \rightarrow s \ell^+ \ell^-$ modes. 
Precision measurements of branching fractions and the lepton forward-backward 
asymmetries in $B \rightarrow X_s \mu^+ \mu^-$,  
$B \rightarrow K^+ \mu^+ \mu^-$ and $B \rightarrow K^{*0} \mu^+ \mu^-$ 
can be achieved in hadron colliders. A super $B$ factory with an annual 
luminosity of $\rm 10^{36} \rm cm^{-2} s^{-1}$, however, is competitive in 
$\mu^+ \mu^-$ final states and, in addition, can measure these quantities
in $e^+ e^-$ final states. Such a machine would also allow for
precise measurements of $B \rightarrow \rho (\omega) \gamma$,
and yield an observation of $B \rightarrow X_d \gamma$. 
Furthermore, one would have a unique opportunity to detect the
$B \rightarrow X_s \nu \bar \nu$ and  $B \rightarrow K^+ (K^{*0}) \nu \bar \nu$
modes and measure their properties, since due to 
$q\bar q$ continuum and $B \bar B$ backgrounds
these rare $B$ decays are not accessible in hadron machines.
However, because of the two escaping $\nu's$ hermiticity of the detector
is a key issue. Since the acceptance of the super $B$-factory
detector is likely to be similar to that of BABAR, one might consider 
of adding a layer of scintillators before the focusing quadrupoles.  


%
%

%
%

\begin{acknowledgments}
This work has been supported by NFR.
\end{acknowledgments}


\bibliography{p3-41-ge}

\end{document}